\documentclass[
 reprint,
superscriptaddress,
 amsmath,amssymb,
 aps,
pra,
]{revtex4-2}

\usepackage{graphicx}
\usepackage{dcolumn}
\usepackage{bm}
\usepackage{hyperref}
\usepackage[utf8]{inputenc}
\usepackage[T1]{fontenc}
\usepackage{booktabs, array, mathptmx, float, tabularx, booktabs, lipsum, amsmath,multirow}
\usepackage{siunitx, xcolor}
\usepackage[version=4]{mhchem}
\usepackage{color}
\usepackage{setspace}
\usepackage{subfigure}
\usepackage[section]{placeins}

\begin{document}
\begin{spacing}{2.0}

\title{Polarized $\gamma$-photon beams produced by collision of two ultrarelativistic electron beams}

\affiliation{Department of Physics and Beijing Key Laboratory of Opto-electronic Functional Materials and Micro-nano Devices, Renmin University of China, Beijing 100872, China}
\affiliation{Key Laboratory of Quantum State Construction and Manipulation (Ministry of Education), Renmin University of China, Beijing, 100872, China}
\affiliation{IFSA Collaborative Innovation Center, Shanghai Jiao Tong University, Shanghai 200240, China}

\author{Zhe Gao}
\affiliation{Department of Physics and Beijing Key Laboratory of Opto-electronic Functional Materials and Micro-nano Devices, Renmin University of China, Beijing 100872, China}
\affiliation{Key Laboratory of Quantum State Construction and Manipulation (Ministry of Education), Renmin University of China, Beijing, 100872, China}

\author{Wei-Min Wang}
\email{weiminwang1@ruc.edu.cn}
\affiliation{Department of Physics and Beijing Key Laboratory of Opto-electronic Functional Materials and Micro-nano Devices, Renmin University of China, Beijing 100872, China}
\affiliation{Key Laboratory of Quantum State Construction and Manipulation (Ministry of Education), Renmin University of China, Beijing, 100872, China}
\affiliation{IFSA Collaborative Innovation Center, Shanghai Jiao Tong University, Shanghai 200240, China}


\begin{abstract}
Many studies have shown that high-energy $\gamma$-photon beams can be efficiently generated via nonlinear Compton scattering driven by laser pulses with intensities $> 10^{22}\rm{W/cm^2}$ recently available in laboratories. Here, we propose a laserless scheme to efficiently generate high-energy polarized $\gamma$-photon beams by collision of two ultrarelativistic electron beams. The self-generated field of a dense driving electron beam provides the strong deflection field for the other ultrarelativistic seeding electron beam. A QED Monte Carlo code based on the locally constant field approximation is employed to simulate the collision process, and the polarization properties of produced $\gamma$ photons are investigated. The simulation results and theoretical analysis indicate that the photon polarization, including both linear and circular polarizations, can be tuned by changing the initial polarization of the seeding beam. If an unpolarized seeding beam is used, linearly polarized photons with an average polarization of 55\% can be obtained. If the seeding beam is transversely (longitudinally) polarized, the linear (circular) polarization of photons above 3 GeV can reach 90\% (67\%), which is favorable for highly polarized, high-energy $\gamma$ photon sources.
\end{abstract}

\maketitle


\section{Introduction}

Polarized $\gamma$-photon beams play an important role in astrophysics \cite{doi:10.1126/science.1200848}, nuclear physics \cite{RevModPhys.77.1131}, high-energy physics \cite{MOORTGATPICK2008131,PhysRevLett.118.204801}, and plasma physics\cite{10.1063/5.0078961}. For example, in high-energy physics, the linear polarized photon beam can be used to distinguish Delbrück scattering from other elastic scattering processes \cite{PhysRevLett.118.204801} and achieve a more accurate measurement of Delbrück scattering. Polarized electrons and positrons can be used to detect nuclear structures accurately\cite{Subashiev:1998me} and verify standard models \cite{MOORTGATPICK2008131} in future electron-positron colliders, where traditionally the circularly polarized $\gamma$-photon beams are usually needed to generate these polarized positrons via the Bethe-Heilter process. 

Polarized $\gamma$ photons can be obtained through Compton scattering \cite{PhysRevLett.96.114801,PhysRevLett.100.210801,PhysRevSTAB.18.110701} and Bremsstrahlung \cite{PhysRev.114.887,PhysRevLett.116.214801}. Unpolarized electrons can radiate polarized photons by linear Compton scattering. The photon polarization is determined by the driving laser polarization because the radiation formation length is longer than the laser wavelength \cite{Tang_2020}. However, the photon polarization imprinted from the laser field could not work in the nonlinear regime. For Bremsstrahlung, the polarization can be transferred from electrons to photons, but the incident electron density should be low enough to avoid the crystal being damaged \cite{biryukov2013crystal}.

With the rapid development of the ultra-short, ultra-intense laser technology, peak intensities of $10^{22}$ $\mathrm{W/cm^2}$, and even $10^{23}$ $\mathrm{W/cm^2}$ can be delivered from PW-level laser facilities \cite{danson_hillier_hopps_neely_2015,Yoon:21}. By use of such laser pulses, many schemes are proposed to generate high-energy $\gamma$-photons with brightness \cite{wang2018collimated,zhu2020extremely} or certain polarization \cite{tang2020highly,PhysRevLett.124.014801,doi:10.1063/5.0007734}. Tang et al. \cite{tang2020highly} showed that circularly polarized photons with a polarization of 78\% or linearly polarized photons with a polarization of 91\% can be obtained by using the interaction of circularly polarized or linearly polarized lasers with electrons in the weakly nonlinear Compton scattering regime. Li et al. \cite{PhysRevLett.124.014801} found that polarized $\gamma$ photons can be emitted by ultrarelativistic pre-polarized electrons through nonlinear Compton scattering, where the polarization of high-energy photons is determined by the electron polarization. Xue et al. \cite{doi:10.1063/5.0007734} proposed to generate polarized $\gamma$ photons by intense laser interaction with plasmas, and the photon brilliance can reach $10^{21}$ photons $\mathrm{/(s\  mm^2\ mrad^2\ 0.1\%\ BW)}$ .

Different from the laser-electron collisions or laser-plasma interactions, here we propose a laserless scheme to generate polarized $\gamma$-photon beams via the collision of two ultrarelativistic electron beams. In the proposed scheme, one electron beam is the driving beam, and the other is the seeding beam. The driving beam can provide a strong self-generated unipolar field, which plays the similar role to the laser field or the plasma response field in the conventional nonlinear Compton scattering scheme. The self-generated azimuthal magnetic field is perpendicular to the radial electric field, under which the seeding beam can experience transverse Lorentz force and strongly emit $\gamma$ photons. Our simulation results show if an unpolarized seeding beam is used, linearly polarized photon beams with an average polarization of 55\% can be generated. Because the field experienced by the seeding beam is unipolar and approximately along one direction, one can control the initial polarization of the seeding beam to adjust the photon polarization to be either linear or circular. The unipolar field is important, otherwise, different field directions lead to photon polarization in different directions, which is unfavorable for high polarization of photons. 

\begin{figure}\includegraphics[width=0.5\textwidth]{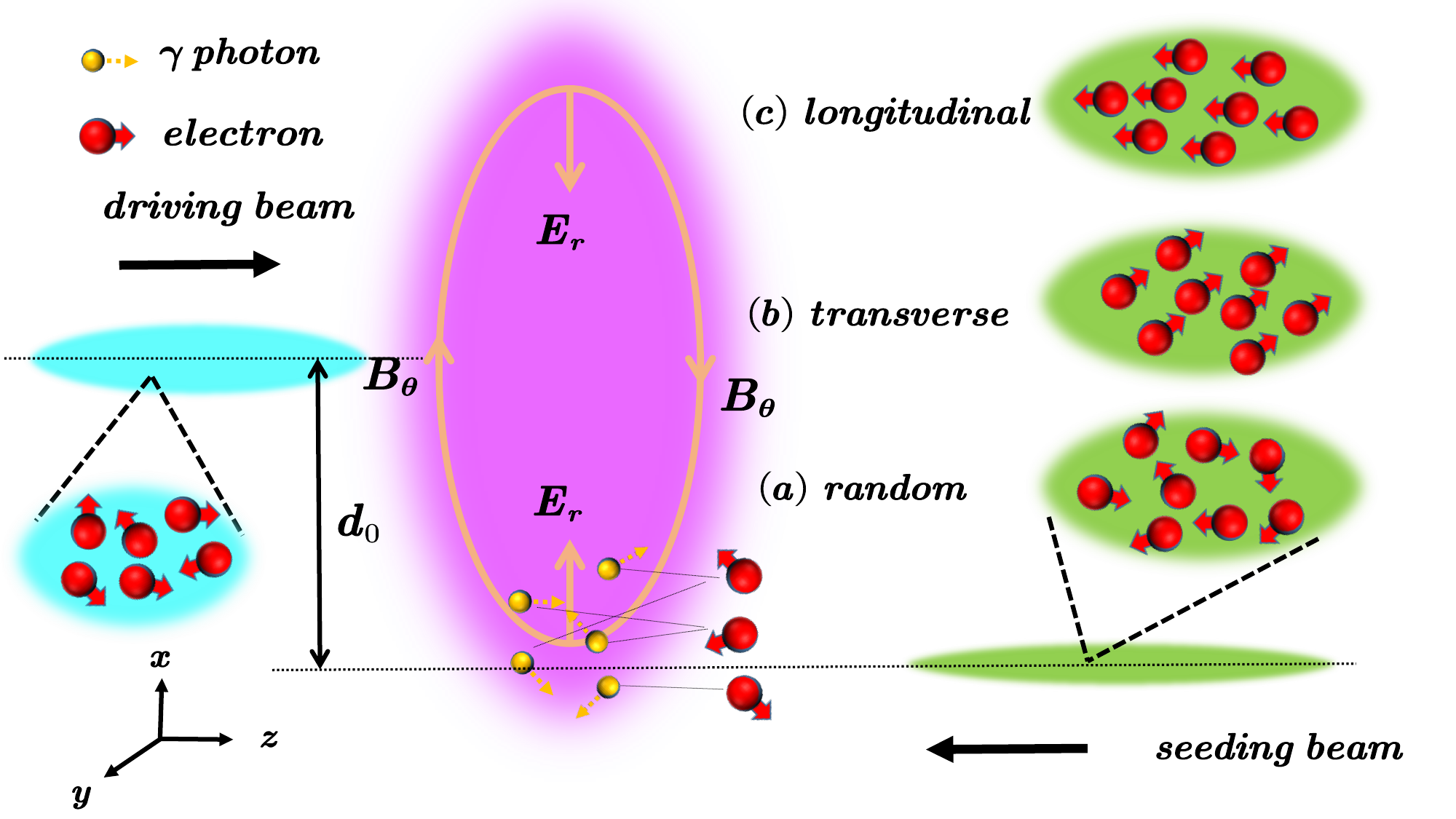}
\caption{\label{fig1} Interaction scenario of an ultrarelativistic seeding electron beam head-on colliding with a dense driving electron beam. The driving beam provides a super-strong self-generated field. In such a transverse field the seeding beam emits polarized $\gamma$ photons, whose emission direction is approximately along the propagation direction of the seeding beam.}
\end{figure}

\section{theoretical model}
We take the charge density of the driving beam as a Gaussian distribution, i.e., $\rho =\rho _0 e^{-\frac{r^2}{2\sigma _{d}^{2}}}e^{-\frac{( z-z_0-vt) ^2}{2l_{d}^{2}}}$, where $\rho _0=\frac{Q_d}{( 2\pi)^{\frac{3}{2}}\sigma _{d}^{2}l_d}$ is the maximum value of the charge density, $Q_d$, $l_d$, and $\sigma_d$ are the total charge, bunch length and bunch width, respectively, $z_0$ is the initial center position of the driving beam and $r=\sqrt{x^2+y^2}$. In the ultrarelativistic case, the self-generated field of the driving beam is approximately $E_r\left( r,z,t \right) \approx B_{\theta}\left( r,z,t \right) \approx 4\pi \rho _0\frac{\sigma _{d}^{2}}{r}\left( 1-e^{-\frac{r^2}{2\sigma _{d}^{2}}} \right) e^{-\frac{\left( z-z_0-vt \right) ^2}{2l_{d}^{2}}}$  \cite{NOBLE1987427,PhysRevLett.126.064801}. There is a maximum field strength $B_\theta^{\max}$ at $r\approx 1.585\sigma _d$ and $z=z_0+vt$, where
\begin{equation}\label{eq:one}
	B_{\theta}^{\max}\approx 1.08\times 10^4\times \frac{Q_d\left( \mathrm{nC} \right)}{\sigma _d\left( \mathrm{\mu m} \right) \times l_d\left( \mathrm{\mu m} \right)}.
\end{equation}
The longitudinal field components $E_z\left( r,z,t \right)$ and $B_z\left( r,z,t \right)$ can be negligible. The maximum value of the QED parameter is about
\begin{equation}\label{eq:two}
	\chi _{e}^{\max}\approx 0.0096\times \frac{Q_d\left( \mathrm{nC} \right) \times \varepsilon _{s0}\left( \mathrm{GeV} \right)}{\sigma _d\left( \mathrm{\mu m} \right) \times l_d\left( \mathrm{\mu m} \right)}, 
\end{equation}
where $\varepsilon_{s0}$ is the initial electron energy of the seeding beam.

In this article, a Monte Carlo algorithm is used to simulate the polarized photon emission \cite{PhysRevSTAB.14.054401,RIDGERS2014273,PhysRevE.92.023305}. The spin-flip of electrons due to radiative polarization and the determination of photon polarization are included \cite{PhysRevLett.124.014801,katkov1998electromagnetic,li2020production}. This code has been used in the previous work to investigate the polarized pair production \cite{PhysRevResearch.3.033245}. We use the spin- and polarization-resolved photon radiation probability derived from the Baier-Katkov method
\begin{equation}\label{eq:three}
	\begin{aligned}
	\frac{d^2W_{\rm rad}}{dudt}&=\frac{\alpha m^2c^4}{4\sqrt{3}\pi \hbar \varepsilon _e}\{ \frac{u^2-2u+2}{1-u}K_{\frac{2}{3}}( y )
	\\
	&-{\rm Int}K_{\frac{1}{3}}( y ) -uK_{\frac{1}{3}}( y ) ( \vec{S}_i\cdot \vec{e}_2 ) 
	\\
	&+\frac{u}{1-u}K_{\frac{1}{3}}( y ) ( \vec{S}_i\cdot \vec{e}_1 ) \xi _1
	\\
	&+\left[ \frac{2u-u^2}{1-u}K_{\frac{2}{3}}( y ) -u{\rm Int}K_{\frac{1}{3}}( y ) \right] ( \vec{S}_i\cdot \vec{e}_v ) \xi _2
	\\
	&+\left. \left[ K_{\frac{2}{3}}( y ) -\frac{u}{1-u}K_{\frac{1}{3}}( y ) ( \vec{S}_i\cdot\vec{e}_2)\right]\xi_3\} \right.
	,
	\end{aligned}
\end{equation}
where $K_{\nu}\left( y \right)$ is the modified Bessel function of the second kind of the $\nu$-th order, $y=\frac{2u}{3\left( 1-u \right) \chi _e}$, $u=\frac{\varepsilon _{\gamma}}{\varepsilon _e}$ is the ratio of the photon energy to the electron energy, $\alpha \approx \frac{1}{137}$ is the fine structure constant, and $\chi _e=\frac{e\hbar}{m_{e}^{3}c^4}\left| F_{\mu \nu}p^{\nu} \right|$ is the nonlinear quantum parameter. Physical variables $e$, $c$, $m_e$, $\hbar$, $F_{\mu \nu}$, and $p^\nu$ are the electron charge, the speed of light, the electron mass, the reduced Planck constant, the electromagnetic field tensor, and the electron four-dimensional momentum, respectively. The normalized vectors $ \vec{S}_i$ and $\vec{S}_f$ are spin vectors of electrons before and after radiation. Since we are mainly concerned with photon polarization, here we have summed the final states of electrons. The formula is based on the locally constant field approximation (LCFA), where the external field remains almost constant when a photon is emitted or the photon emission period is much shorter than the field duration. If the external field satisfies $a_0=\frac{\left| e \right|E}{m_ec\omega _0}\gg 1$, it is generally considered to satisfy the LCFA \cite{ritus1985quantum,di2018implementing,di2019improved,ilderton2019extended,podszus2019high}, where $a_0$  is the normalized field strength, $\omega_0$ is the laser frequency, and $E$ is the deflection field strength. The self-generated radial field $E_r\left( r,z,t \right) $ formed by the driving beam can be considered as a half-cycle laser field with a wavelength of 4$l_d$. The classical electron dynamics are described by the Newton Lorentz force and the TBMT theory is used to describe the electron spin precession \cite{PhysRevLett.2.435}.

For the polarized photon emission, the normalized Stokes vector $\vec{\xi}=\left( \xi _1,\xi _2,\xi _3 \right) $ is used to describe the photon polarization. The Stokes vector depends on the specific coordinate system, and we select $\left( \vec{e}_1,\vec{e}_2,\vec{e}_v \right)$, where $\vec{e}_1$ is the direction of the transverse acceleration of the electron, $\vec{e}_v$ is the velocity direction of the electron, and $\vec{e}_2=\vec{e}_v\times \vec{e}_1$. The Stokes component $\xi _3=+1\left( -1 \right)$ means that the photon is linearly polarized along the direction $\vec{e}_1$($\vec{e}_2$), and $\xi _2=+1\left( -1 \right)$ means that the photon is right-handed (left-handed) circularly polarized. In the ultrarelativistic case, the direction of the photon emission is close to the initial electron velocity direction.

\section{Simulation results and analysis}

\subsection{Simulation parameters}

In the typical simulation case, we take the driving electron beam with the charge $Q_d=4$ $\mathrm{nC}$, the length $l_d=0.5$ $\mathrm{\mu m}$, the width $\sigma _d=1.0$ $\mathrm{\mu m}$, and the initial energy $\varepsilon _{d0}=10$ $\mathrm{GeV}$. The seeding beam has the charge $Q_s=0.128$ pC, the length $l_s=1.0$ $\mathrm{\mu}$m, the width $\sigma _s=0.5$ $\mathrm{\mu m}$, and the initial energy $\varepsilon _{s0}=5$ $\mathrm{GeV}$. The energy spread of the seeding beam is $\varDelta \varepsilon _{s0}/\varepsilon _{s0}=0.01$. The self-generated field of the seeding beam is $10^{-5}$ orders of magnitude lower than the driving beam and can be ignored. Such a driving beam and seeding beam can be provided by FACET II \cite{PhysRevAccelBeams.22.101301} and future laser wakefield accelerators \cite{PhysRevLett.121.264801}. At the initial time, the central positions of the driving beam (along the $+z$ axis) and the seeding beam (along the $-z$ axis) are (x, y, z) = $\left( 0, 0, -2~ \mathrm{\mu m} \right) $ and $\left( d_0, 0, 2~ \mathrm{\mu m} \right) $, respectively, where $d_0$ is the impact parameter between two electron beams, as shown in Fig.~\ref{fig1}.

In our simulation, we set $\sigma _s=0.5$ $\mathrm{\mu m}$ and $d_0=-2$ $\mathrm{\mu m}$ to obtain efficient photon generation. The maximum value of $\chi _e$ in the interaction is 0.38, so the QED effect needs to be taken into account. In all the simulations $a_0\gg 1$, ensuring that the LCFA holds.

\subsection{Simulation results}

\begin{figure}
	\includegraphics[width=0.5\textwidth]{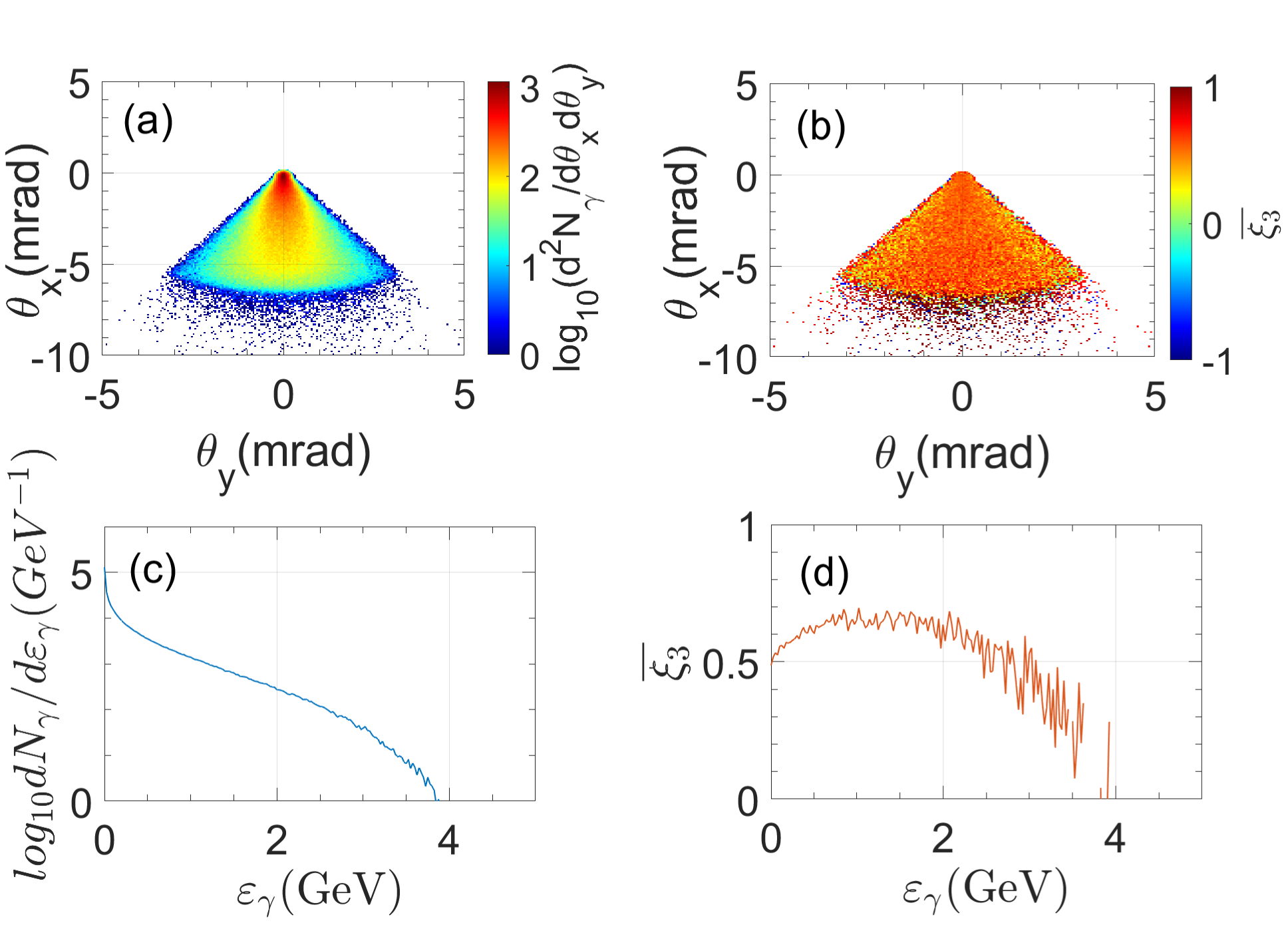}
	\caption{\label{fig2}(a) Photon number density $\log _{10}\left( d^2N_{\gamma}/d\theta _xd\theta _y \right) $ and (b) photon polarization $\overline{\xi _3}$ versus deflection angles $\theta _x$ and $\theta _y$, where $\theta _x\approx \frac{p_x}{\left| p_z \right|}$ and $\theta _y\approx \frac{p_y}{\left| p_z \right|}$. (c) Photon spectrum $\log _{10}dN_{\gamma}/d\varepsilon _{\gamma}$ and (d) photon polarization $\overline{\xi _3}$ versus photon energy $\varepsilon _{\gamma}$.}
\end{figure}

We show the angular distributions, spectrum, and polarization of the generated photons in Fig.~\ref{fig2}. Figure~\ref{fig2} (a) plots the photon angular distribution with respect to $\theta _x$ and $\theta _y$, where $\theta _x\approx \frac{p_x}{\left| p_z \right|}$ and $\theta _y\approx \frac{p_y}{\left| p_z \right|}$. The photons are mainly distributed in the small regions of -6 mrad $<\theta _x<$ 0 and -0.3 mrad $<\theta _y<$ 0.3 mrad. The total number of photons $N_{\gamma}$ is about $4.1\times 10^5$ and the energy conversion efficiency is up to $\eta =\sum_{N_{\gamma}}{\varepsilon _{\gamma}}/\sum_{N_s}{\varepsilon _{s0}}\approx 3.0\%
$ [see Fig.~\ref{fig3}(c)]. The generated photons are mainly distributed along the $-x$ direction because the Lorentz force deflects the seeding beam along the $-x$ direction. If $d_0>0$, the photons will be distributed in a positive x direction, because the self-generated field polarity is reversed. Figure~\ref{fig2}(c) shows the distribution of photon number with photon energy and the cutoff energy of the produced photon approaches 4 GeV.

The produced photons are linearly polarized and the average polarization degree is about 55\% [see Fig.~\ref{fig2}(d) and Fig.~\ref{fig3}(b)]. In our case $\overline{\xi _1}\approx 0$ , $\overline{\xi _2}\approx 0$ , and the average polarization degree is defined as $P_{\gamma}=\sqrt{\overline{\xi _1}^2+\overline{\xi _2}^2+\overline{\xi _3}^2}\approx \overline{\xi _3}$. According to Eq.~(\ref{eq:three}), the average linear polarization degree of photons can be expressed as 
\begin{equation}\label{eq:four}
	\overline{\xi_3}=\frac{ K_{\frac{2}{3}}( y )-\frac{u}{1-u}K_{\frac{1}{3}}( y ) ( \vec{S}_i\cdot\vec{e}_2)}{\frac{u^2-2u+2}{1-u}K_{\frac{2}{3}}( y )-{\rm Int}K_{\frac{1}{3}}( y ) -uK_{\frac{1}{3}}( y ) ( \vec{S}_i\cdot \vec{e}_2 )}.
\end{equation}
In the case of the unpolarized seeding beam, the influences of spin on the polarization degree of photons cancel each other, so one can take $\vec{S}_i\cdot \vec{e}_2=0$. Then, $\overline{\xi _3}$ can be simplified to 
\begin{equation}\label{eq:five}
	\overline{\xi_3}\approx\frac{ K_{\frac{2}{3}}( y )}{\frac{u^2-2u+2}{1-u}K_{\frac{2}{3}}( y )-{\rm Int}K_{\frac{1}{3}}( y )}.
\end{equation}
From Fig.~\ref{fig2} (d), one can see that the linear polarization of the photon increases from 50\% to 69\%  at the energy of 1 GeV and then decreases slowly. When $u\ll1$, ${\rm Int}K_{\frac{1}{3}}(y)\ll K_{\frac{2}{3}}(y)$, and $\overline{\xi _3}\approx \frac{1}{2}$. As u increases, ${\rm Int}K_{\frac{1}{3}}(y)/K_{\frac{2}{3}}(y)$ increases rapidly, resulting in an increase in $\overline{\xi _3}$. According to Eq.~(\ref{eq:five}), when u is larger than 0.7, ${\rm Int}K_{\frac{1}{3}}(y)\approx 0.9K_{\frac{2}{3}}(y)$ and $\overline{\xi _3}\approx \frac{1}{\frac{u^2-2u+2}{1-u}-0.9}=\frac{1-u}{u^2-1.1u+1.1}$, which is very close to our simulation shown in Fig.~\ref{fig2}(d). This figure shows the high-energy photons tend to have low polarization and these photons usually distribute in small $\left| \theta _x \right|$ and $\left| \theta _y \right|$ as observed in Fig.~\ref{fig2}(b). High-energy photons are mainly produced by high-energy electrons distributed in small $\left| \theta _x \right|$ and $\left| \theta _y \right|$. Low-energy photons distributed in large $\left| \theta _x \right|$ and $\left| \theta _y \right|$ have higher degrees of polarization whose parent electrons radiate strongly and lose substantial energies.

\subsection{Influence of beam parameters}

\begin{figure}
	\includegraphics[width=8.6cm]{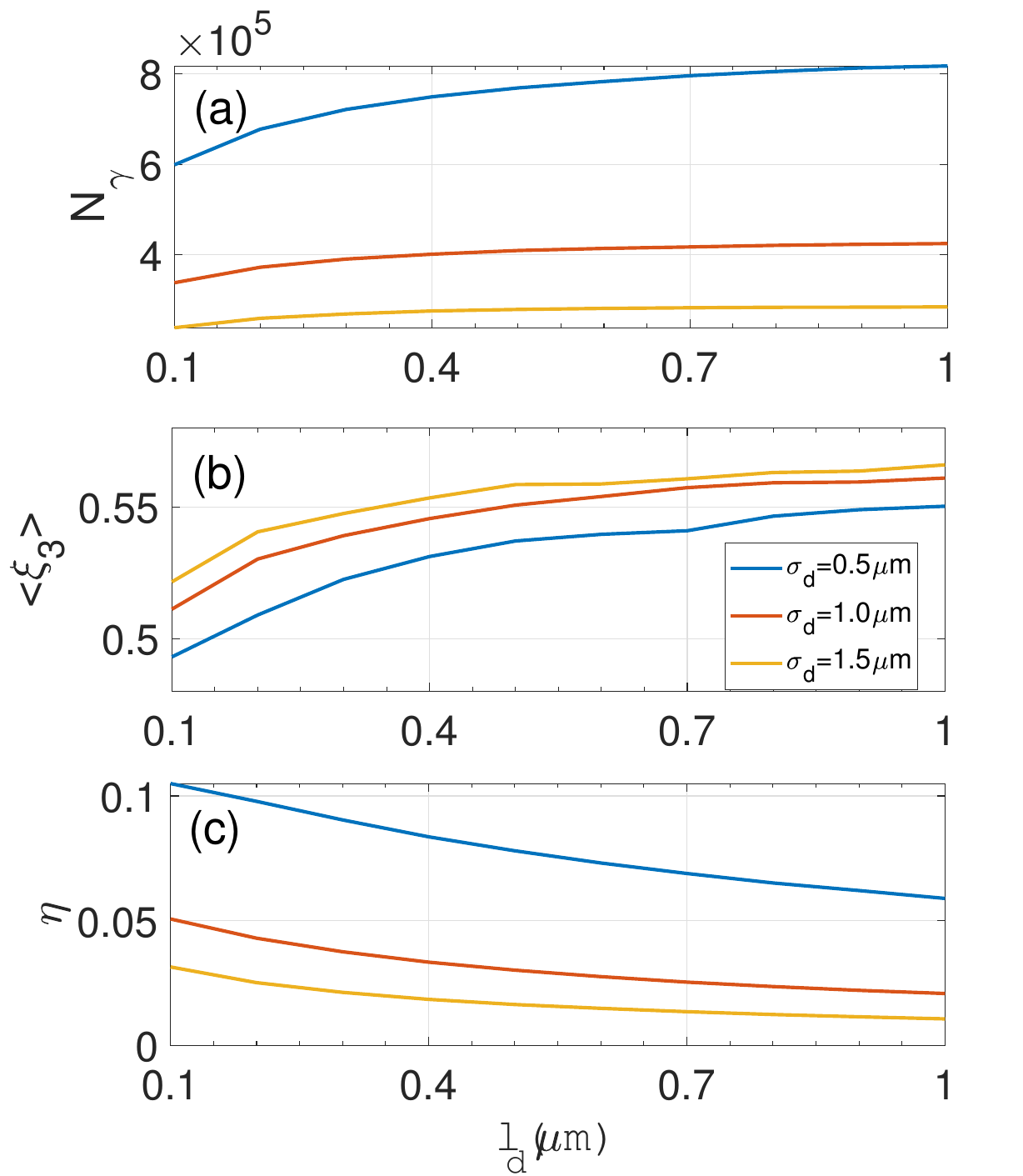}
	\caption{\label{fig3}
		(a) Photon number $N_{\gamma}$, (b) photon polarization $<\xi _3>$ and (c) energy conversion efficiency $\eta$ versus the driving beam length $l_d$ in three cases with the driving beam widths of $\sigma _d=0.5$ $\mathrm{\mu m}$, $1.0$ $\mathrm{\mu m}$ and $1.5$ $\mathrm{\mu m}$, respectively.}
\end{figure}

In Fig.~\ref{fig3}, the influences of the length $l_d$ of the driving beam on the photon number $N_{\gamma}$, photon polarization $<\xi _3>$, and energy conversion efficiency $\eta$ under different beam widths $\sigma _d$ are studied, where the charge of the driving beam $Q_d$ is 4~nC and $\sigma _s/\sigma _d=0.5$. Comparing different lines in Fig.~\ref{fig3}(a) shows that the photon number increases with the decrease of the driving beam width $\sigma_d$ since a smaller $\sigma_d$ causes a larger $\chi _e$[see Eq.(2)] and consequently a higher radiation probability. This figure also displays that the photon number $N_\gamma$ increases with the increase of $l_d$, since $N_{\gamma}\sim \frac{l_d}{c}\int du \frac{d^2W_{\rm rad}}{dudt}$, where $\frac{d^2W_{\rm rad}}{dudt}$ is the probability of radiation per unit of time per unit of energy. The photon number $N_{\gamma} \propto l_d^{1/3}$ at $\chi_e\gg1$ and $N_{\gamma} \propto 1$ at $\chi_e\ll1$, where the probability of radiation $\frac{dW_{\rm rad}}{dt}$ is inversely proportional to $l_d^{2/3}$ and $l_d$ in the two cases \cite{berestetskii1982quantum}, respectively. Note that for the longer $l_d$, the average energy of photons tends to decrease since $\chi_e$ is smaller according to Eq.~(\ref{eq:two}). As shown in Fig.~\ref{fig3}(c), energy conversion $\eta$ decreases with the increase of $l_d$ and $\sigma _d$, because the radiation energy of an electron is mainly determined by $\chi_e$ which is inversely proportional to $l_d$ and $\sigma_d$. Note that the reduction of $\eta$ and the increase of $N_{\gamma}$ with the growing $l_d$ means that the average photon energy and the proportion of high-energy photons is reduced. 

In Fig.~\ref{fig3}(b), one can see that the average polarization of photons increases with the increase of $l_d$ and $\sigma _d$. By summing the photon energy spectrum, one can obtain the expression of the total linear polarization of photons: 
\begin{equation}\label{eq:six}
	<\xi_3>=\frac{6\int\limits_0^{\infty}{dv\frac{\exp \left( -f \right)}{\left( 1+v \right) ^2}fv\left( 4+3v-fv \right)}}{\int\limits_0^{\infty}{dv\frac{\exp \left( -f \right)}{\left( 1+v \right) ^3}fv\left[ 3+29\left( 1+v \right) ^2-3fv\left( 2+v \right) +f^2v^2 \right]}},
\end{equation}
where v is the integral variable, $f=v(3(1+v))^{1/2}/\chi_{e}$, and more details can be found in Appendix A. In Eq.~(\ref{eq:six}), $<\xi_3>$ decreases monotonically with $\chi_e$ (see Fig.~\ref{fig8} in Appendix A). According to Eq.~(\ref{eq:two}), $\chi_e$ is inversely proportional to $l_d$ and $\sigma _d$, so the average polarization of photons increases with the increase of $l_d$ and $\sigma _d$.

\begin{figure}[b]
	\includegraphics[width=0.5\textwidth]{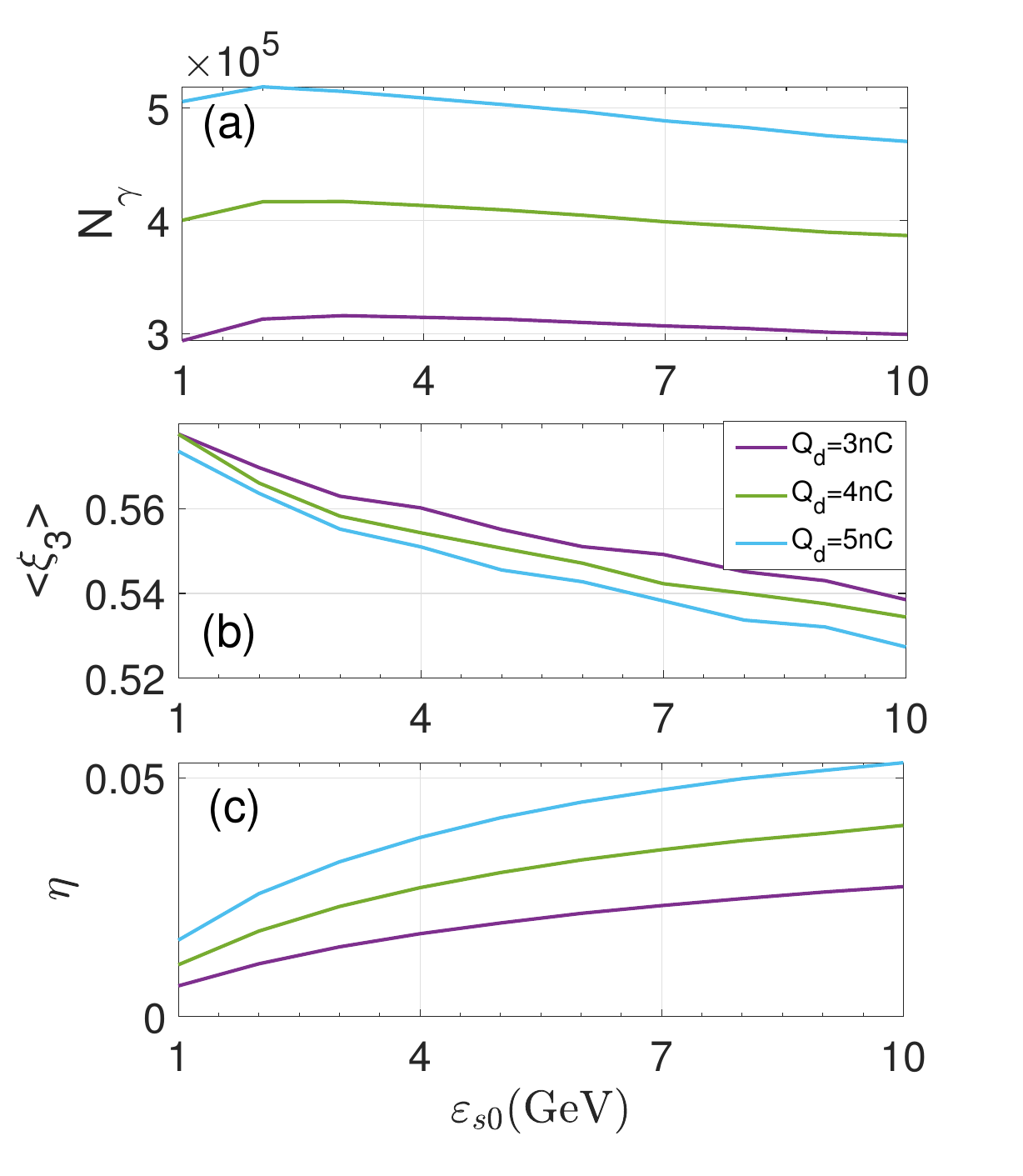}
	\caption{\label{fig4}
		(a) Photon number $N_{\gamma}$, (b) photon polarization $<\xi _3>$ and (c) energy conversion efficiency $\eta$ versus the seeding beam initial energy $\varepsilon _{s0}$ in three cases with the charges of driving beam $Q_d=3$ $\mathrm{nC}$, $4$ $\mathrm{nC}$ and $5$ $\mathrm{nC}$, respectively.}
\end{figure}

Figure~\ref{fig4} shows the influence of the initial energy of the seeding electron beam on the photon number $N_{\gamma}$, photon polarization $<\xi _3>$ and energy conversion efficiency $\eta$, respectively. Increasing the seeding beam energy $\varepsilon_{s0}$ results in the reduction of the photon number and polarization as shown in Figs.~\ref{fig4}(a) and ~\ref{fig4}(b), respectively. We notice that the increase of $\varepsilon_{s0}$ has a slight effect on the photon number, e.g., the number of photons does not change by more than 5\% when $\varepsilon_{s0}$ is increased from 1 GeV to 10 GeV with $Q_d=4$ nC. This can be explained by Eq.~(\ref{eq:three}) that $W_{\rm rad}$ is inversely proportional to the electron energy $\varepsilon_e$, which offsets the effect of the increase of $\chi_e$. Increasing $\varepsilon_{s0}$ leads to the increase in $\chi_e$ and the decrease in $<\xi _3>$, in terms of Eq.~(\ref{eq:six}) and the discussion above. In Fig.~\ref{fig4}(c), it is found that increasing the seeding beam energy can improve the energy conversion efficiency, as well as grow the proportion of high-energy photons. This is because $\eta \propto \int du \varepsilon_{\gamma} \frac{d^2W_{\rm rad}}{dudt}= \int du \frac{\alpha m^2c^4}{4\sqrt{3}\pi \hbar} u \{ \frac{u^2-2u+2}{1-u}K_{\frac{2}{3}}( y )-{\rm Int}K_{\frac{1}{3}}( y )\}=\int du \frac{\alpha m^2c^4}{4\sqrt{3}\pi \hbar} u \{ \frac{u^2}{1-u}K_{\frac{2}{3}}( y )+{\rm Int}K_{\frac{5}{3}}( y )\}$, and $K_{\frac{2}{3}}$ and ${\rm Int}K_{\frac{5}{3}}$ increase with the decrease of y which is inversely proportional to $\varepsilon_{s0}$.

Increasing the charge of the driving beam $Q_d$ also causes the growth of $N_\gamma$, but the decrease of $<\xi _3>$. This is because with the increase of $\chi_e$ as the growing $Q_d$, $N_\gamma$ goes up and $<\xi _3>$ goes down[according to Eq.~(\ref{eq:six})]. One can increase the number of photons and the energy conversion efficiency by increasing $Q_d$ according to Figs.~\ref{fig4}(a) and ~\ref{fig4}(b). In the meanwhile, the increased $Q_d$ brings a slight negative impact on the photon polarization.

\subsection{Improving the photon polarization through the initial polarization of the seeding beam}

\begin{figure}
	\includegraphics[width=8.6cm]{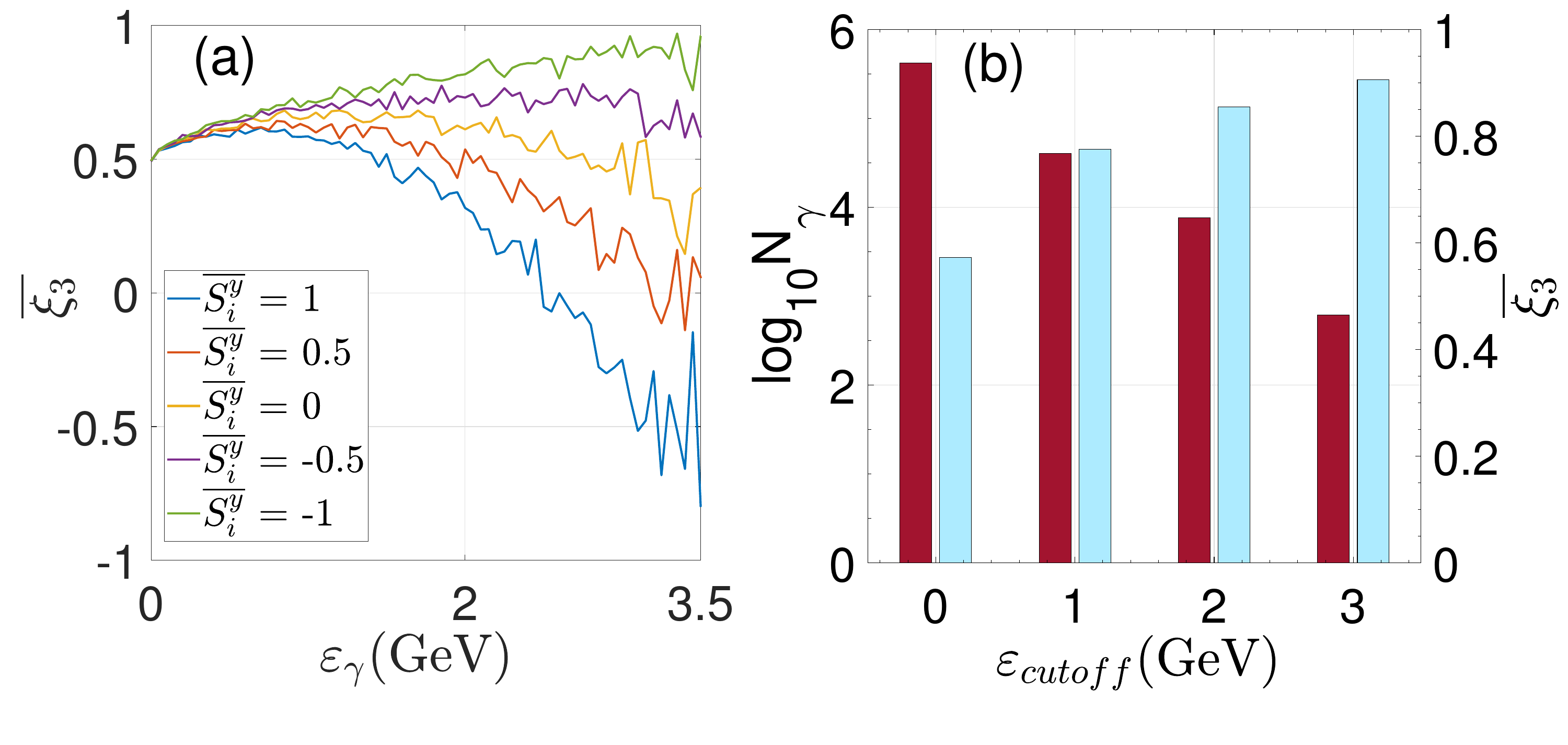}
	\caption{\label{fig5}(a) Photon polarization $\overline{\xi _3}$ versus photon energy $\varepsilon _{\gamma}$ with various initial electron polarization $\overline{S_{i}^{y}}=-1$, $-0.5$, $0$, $0.5$ and $1$. (b) Photon number $N_{\gamma}$(red bar) and average photon polarization $\overline{\xi _3}$ (blue bar) versus the cutoff energy $\varepsilon _{\gamma}$.}
\end{figure}

Above we have taken an initially unpolarized seeding electron beam, and the polarization of the photons in the high-energy tail is relatively low, as shown in Fig.~\ref{fig2} (d). To improve the polarization of high-energy photons, one can use a pre-polarized seeding beam with a polarization along the y-axis(or a transverse polarization), i.e., the spin direction of the electrons is the same as or opposite to the direction of the experienced effective magnetic field in the electron rest frame. The average polarization is given by Eq.~(\ref{eq:four}). The first term: 
\begin{equation}\label{eq:seven}
	\frac{ K_{\frac{2}{3}}( y )}{\frac{u^2-2u+2}{1-u}K_{\frac{2}{3}}( y )-{\rm Int}K_{\frac{1}{3}}( y ) -uK_{\frac{1}{3}}( y ) ( \vec{S}_i\cdot\vec{e}_2)}
\end{equation}
is degenerated to $\frac{ K_{\frac{2}{3}}( y )}{\frac{u^2-2u+2}{1-u}K_{\frac{2}{3}}( y )-{\rm Int}K_{\frac{1}{3}}( y )}$ in the case with an initially unpolarized seeding beam and it decays to 0 as u increases, which leads to low polarization of the high-energy photons. The second term of Eq.~(\ref{eq:four}): 
\begin{equation}\label{eq:eight}
	\frac{ \frac{u}{1-u}K_{\frac{1}{3}}( y ) ( \vec{S}_i\cdot\vec{e}_2)}{\frac{u^2-2u+2}{1-u}K_{\frac{2}{3}}( y )-{\rm Int}K_{\frac{1}{3}}( y ) -uK_{\frac{1}{3}}( y ) ( \vec{S}_i\cdot \vec{e}_2 )}
\end{equation}
determines the polarization of high-energy photons, where $\frac{u}{1-u}K_{\frac{1}{3}}( y )$ increases monotonically with u and takes the maximum value around $u=1$. When $\vec{S}_i\cdot \vec{e}_2<0$, $\overline{\xi _3}$ will increase towards 1 as photon energy increases, and when $\vec{S}_i\cdot \vec{e}_2>0$, $\overline{\xi _3}$ will decrease towards -1, which suggests that a pre-polarized seeding beam can generate highly polarized photons with high energies. 

In Fig.~\ref{fig5}(a), we draw the photon polarization distribution when the initial polarization of the seeding beam $\overline{S_{i}^{y}}=-1$, $-0.5$, $0$, $0.5$ and $1$. The average polarizability of the generated photons is 0.57, 0.56, 0.55, 0.54 and 0.53, respectively. For high-energy photons with energies higher than 3.5 GeV, $\overline{\xi _3}=0.94$, $0.69$, $0.35$, $-0.05$, and $-0.80$, respectively. It is found that when $\overline{S_{i}^{y}}=1$ (the electron spins are in the same direction as the magnetic field), $\overline{\xi _3}$ will first increase from about 0.5 and then decrease to -1 as the photon energy increases. If the electron spins are in the opposite direction to the magnetic field with $\overline{S_{i}^{y}}=-1$, $\overline{\xi _3}$ increases from about 0.5 to 1 as photon energy increases, so the average polarization is higher than the case with $\overline{S_{i}^{y}}=1$. These results agree with Eq.~(\ref{eq:four}) and the corresponding calculation of this equation is shown in Fig.~\ref{fig7}(a).

In Fig.~\ref{fig5}(b), we plot the number and polarizability of high-energy photons at different energy cutoff $\varepsilon_{\rm cutoff}$, where we counted the photons with energy above $\varepsilon_{\rm cutoff}$. With the increase of $\varepsilon_{\rm cutoff}$, the number of photons is decreased and the photon polarization is increased, e.g., 0.15\% of photons have energies over 3 GeV, but their average polarization can reach 90\%. By contrast, the polarization is 43\%, when the seeding beam is unpolarized initially. The number ratio of the photons with energies over 1 GeV and 2 GeV is 9.6\% and 1.8\%, and the corresponding average polarization is 78\% and 86\%, respectively.

\begin{figure}[b]
	\centering
	\includegraphics[width=8.6cm]{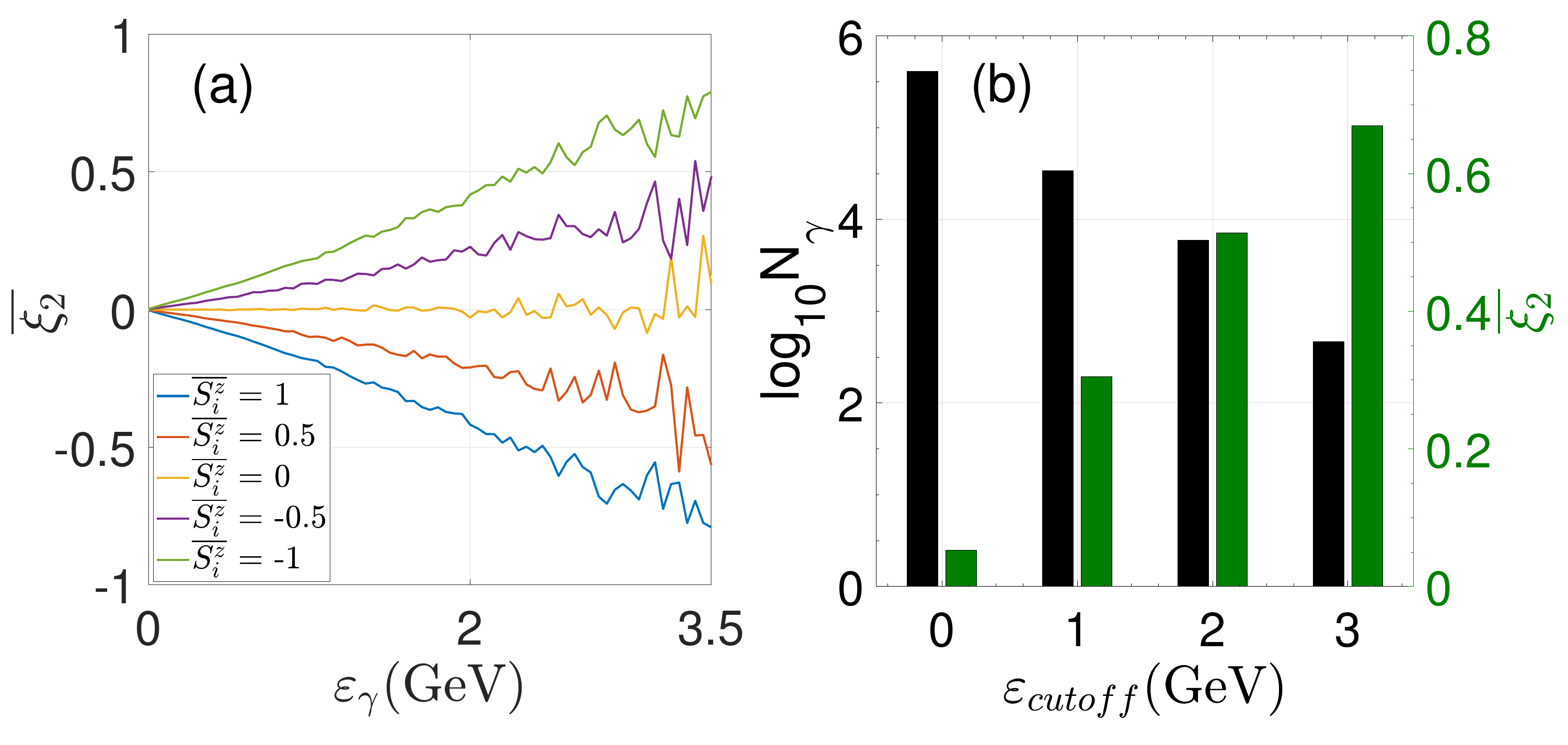}
	\caption{\label{fig6} (a) Photon polarization $\overline{\xi _2}$ versus photon energy $\varepsilon _{\gamma}$ under various initial electron polarization $\overline{S_{i}^{z}}=-1$, $-0.5$, $0$, $0.5$ and $1$. (b) Photon number $N_{\gamma}$(black bar) and average photon polarization $\overline{\xi _2}$(green bar) versus the cutoff energy $\varepsilon _{\gamma}$.}
	
\end{figure}

Next, we proceed to study the generation of circularly polarized $\gamma$ photons, which is related to the longitudinal polarization of the seeding beam. The circular polarization of photons can be written as 
\begin{equation}\label{eq:nine}
	\overline{\xi_2}=( \vec{S}_i\cdot \vec{e}_v )\frac{ \frac{2u-u^2}{1-u}K_{\frac{2}{3}}( y )-u{\rm Int}K_{\frac{1}{3}}( y )}{\frac{u^2-2u+2}{1-u}K_{\frac{2}{3}}( y )-{\rm Int}K_{\frac{1}{3}}( y )},
\end{equation} 
and it is shown that a longitudinally pre-polarized seeding beam with $\vec{S}_i\cdot \vec{e}_v \neq 0$  is necessary to achieve nonzero $\overline{\xi_2}$. Figure~\ref{fig6}(a) shows the influence of the initial polarization of the seeding beam on the circular polarization of the generated $\gamma$ photons. If an initially unpolarized electron beam is used, no circularly polarized photons can be produced, as observed in the line with $\overline{S_{i}^{z}}=0$ in Fig.~\ref{fig6}(a). Left-handed circularly polarized photons can be produced when the spin direction of the electron is along its movement direction. Right-handed circularly polarized photons can be produced when these two directions are opposite. With an unpolarized beam, the numbers of the left-handed and right-handed photons are the same, and they can be summed to linear polarized photons. If the seeding beam has initially longitudinal polarization with $\overline{S_{i}^{z}}=\pm 1$ and $\pm0.5$, circularly-polarized photons can be obtained and the circular polarization grows with the increasing $|\overline{S_{i}^{z}}|$ [see Fig.~\ref{fig6}(a)], where the average circular polarization is  5.3\%, with $\overline{S_{i}^{z}}=-1$. The circular polarization grows with the increase of the photon energy, as shown in Fig.~\ref{fig6}(a). In Fig.~\ref{fig6}(b), we plot the circular polarization of photons in different photon energy cutoffs when $\overline{S_{i}^{z}}=-1$ is taken. As $\varepsilon_{\rm cutoff}$ is increased from 1 GeV to  2 GeV and 3 GeV, the average circular polarization goes up from 30\% to 51\% and 67\%, respectively. This suggests that the scheme with a longitudinally pre-polarized seeding beam is favorable for generating high-energy circularly-polarized photons with sufficiently high polarization.

\begin{figure}\centering\includegraphics[width=8.6cm]{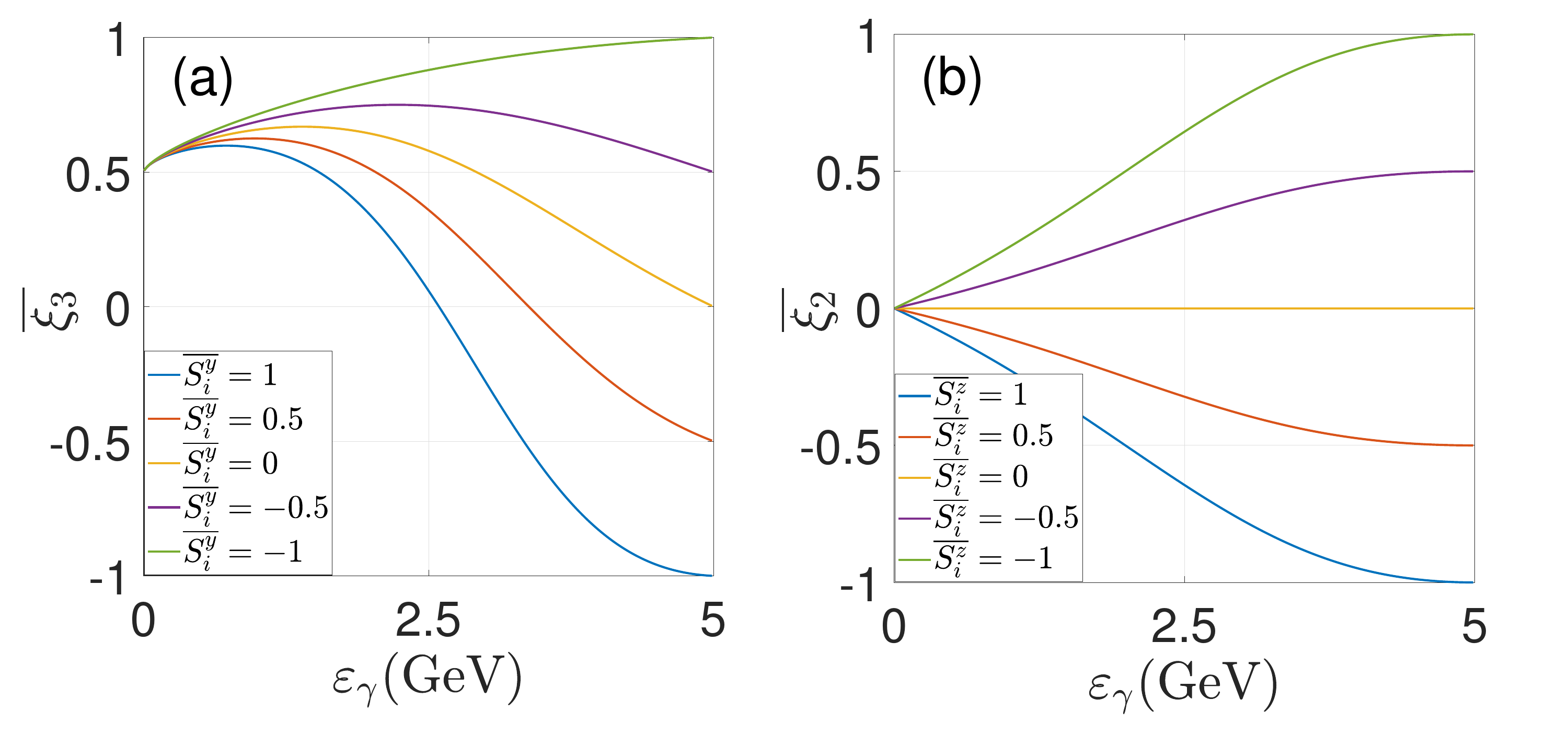}
	\caption{\label{fig7} (a) Linear photon polarization $\overline{\xi _3}$ versus photon energy $\varepsilon _{\gamma}$ under various initial electron polarization of $\overline{S_{i}^{y}}=-1$, $-0.5$, $0$, $0.5$ and $1$. (b) Circular photon polarization $\overline{\xi _3}$ versus photon energy $\varepsilon _{\gamma}$ under various initial electron polarization of $\overline{S_{i}^{z}}=-1$, $-0.5$, $0$, $0.5$ and $1$, repectively. Here, $\chi_e$ = 0.38 is taken, which is close the typical value obtained in our simulation. }
\end{figure}

We calculate the photon polarization with different initial polarizations of the seeding beam according to Eqs.~(\ref{eq:four}) and (\ref{eq:nine}) and present in Fig.~\ref{fig7}. Figure~\ref{fig7}(a) shows the linear photon polarization $\overline{\xi _3}$ with different initial electron polarization and Fig.~\ref{fig7}(b) shows the circular photon polarization. One can see that the theoretical results are close to the simulation results present in Figs.~\ref{fig5}(a) and ~\ref{fig6}(a) with different initial electron polarization. Note that $\overline{\xi_2}$ appears symmetric distributions $\overline{S_{i}^{z}}=1$ and $\overline{S_{i}^{z}}=-1$ as well as $\overline{S_{i}^{z}}=0.5$ and $\overline{S_{i}^{z}}=-0.5$ since $\overline{\xi_2}\propto \vec{S}_i\cdot \vec{e}_v$ as shown in Eq.~(\ref{eq:nine}).

\section{Conclusion}

Through a series of Monte Carlo simulations, it is found that a linearly polarized photon beam with an average polarization of 55\% can be obtained by the collision of two unpolarized ultrarelativistic electron beams. Increasing the charge of the driving beam and decreasing the width of the driving beam can significantly increase the photon yield, with little effect on the photon polarization. When a pre-polarized seeding beam is used, the photons with high energies could have a high polarization. For example, if the transverse polarization of the seeding beam is $\overline{S_{i}^{y}}=-1$, the average polarization of high-energy photons with energies greater than 3GeV can reach 90\%. Circularly polarized photons can be obtained by using an initially longitudinally polarized seeding beam and the average circular polarization of the photons with energy $>$3 GeV can reach 67\%. 

\begin{acknowledgments}
This work was supported by the Strategic Priority Research Program of Chinese Academy of Sciences (Grant No. XDA25050300), the National Key R\&D Program of China (Grant No. 2018YFA0404801), and the Fundamental Research Funds for the Central Universities, the Research Funds of Renmin University of China (20XNLG01). Computational resources have been provided by the Physical Laboratory of High Performance Computing at Renmin University of China.
\end{acknowledgments}

\appendix
\section{}

We derive the dependency of $\xi_3$ on $\chi_e$. The radiation intensity matrix can be expressed as
\begin{equation}\label{eq:ten}
	\begin{aligned}
		dI_{12}&=dI_{21}=0
		\\
		dI_{11}&=\frac{e^2m^2\zeta d\zeta}{2\sqrt{3}\pi \hbar ^2\left( 1+\eta \right) ^3}\left\{ \int_{2\zeta /3\chi_e}^{\infty}{K_{5/3}\left( y \right) dy} \right. 
		\\
		&\left. +\left[ 1+\frac{\zeta ^2}{1+\zeta} \right] K_{2/3}\left( \frac{2\zeta}{3\chi_e} \right) \right\} 
		\\
		dI_{22}&=\frac{e^2m^2\zeta d\zeta}{2\sqrt{3}\pi \hbar ^2\left( 1+\zeta \right) ^3}\left\{ \int_{2\zeta /3\chi_e}^{\infty}{K_{5/3}\left( y \right) dy} \right. 
		\\
		&\left. +\left[ -1+\frac{\zeta ^2}{1+\zeta} \right] K_{2/3}\left( \frac{2\zeta}{3\chi_e} \right) \right\} ,
	\end{aligned}
\end{equation}
where $\zeta =\frac{\hbar \omega}{\varepsilon _e-\hbar \omega}$.
Then we integrate over $\zeta$ and get the formula
\begin{equation}\label{eq:eleven}
	\begin{aligned}
		I_e&=I_{11}+I_{22}
		\\
		&=\frac{e^2m^2}{\sqrt{3}\pi \hbar ^2}\int_0^{\infty}{\frac{\zeta d\zeta}{\left( 1+\zeta \right) ^3}}\left\{ \int_{2\zeta /3\chi_e}^{\infty}{K_{5/3}\left( y \right) dy} \right. 
		\\
		&\left. +\frac{\zeta ^2}{1+\zeta}K_{2/3}\left( \frac{2\zeta}{3\chi_e} \right) \right\} 
		\\
		&=\frac{e^2m^2}{3\sqrt{3}\pi \hbar ^2}\int_0^{\infty}{\frac{\zeta \left( 4\zeta ^2+5\zeta +4 \right)}{\left( 1+\zeta \right) ^4}K_{2/3}\left( \frac{2\zeta}{3\chi_e} \right) d\zeta}
	\end{aligned}
\end{equation}
\begin{equation}\label{eq:twelve}
I_-=I_{11}-I_{22}=\frac{e^2m^2}{\sqrt{3}\pi \hbar ^2}\int_0^{\infty}{\frac{\zeta d\zeta}{\left( 1+\zeta \right) ^3}}K_{2/3}\left( \frac{2\zeta}{3\chi_e} \right) 
\end{equation}
Then, we use the integral representation of the McDonald function
\begin{equation}\label{eq:thirteen}
	\int_{-\infty}^{\infty}{x\sin \left( bx+ax^3 \right) dx}=\frac{2}{3\sqrt{3}}\frac{b}{a}K_{2/3}\left( \sigma \right) ,
\end{equation}
where $\sigma=(2/3\sqrt{3})(b^{3/2}/a^{1/2})$ and obtain 
\begin{equation}\label{eq:fourteen}
	I_e=\frac{e^2m^2}{6\pi i\hbar ^2}\int_0^{\infty}{\frac{4\zeta ^3+5\zeta ^2+4\zeta}{\left( 1+\zeta \right) ^4}}\int_{-\infty}^{\infty}{d\tau \tau e^{i\zeta \left( \tau +\tau ^3/3 \right) /\chi_e}}
\end{equation}
\begin{equation}\label{eq:fifteen}
	I_-=\frac{e^2m^2}{2i\pi \hbar ^2}\int_0^{\infty}{\frac{\zeta}{\left( 1+\zeta \right) ^3}}\int_{-\infty}^{\infty}{d\tau \tau e^{i\zeta \left( \tau +\tau ^3/3 \right) /\chi_e}}.
\end{equation}
\begin{figure}
	\centering
	\includegraphics[width=8.6cm]{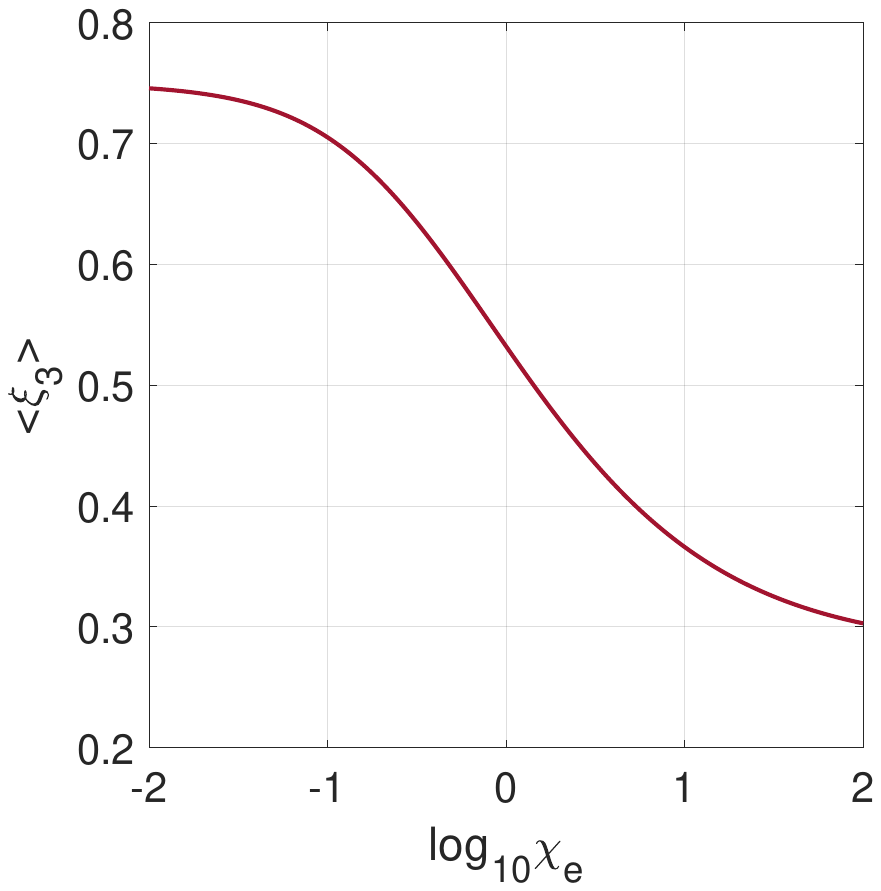}
	\caption{\label{fig8} Photon polarization $<\xi _3>$ versus QED parameter $\chi_e$.}
	
\end{figure} 
Changing the variable $v=\zeta \left( 1+\tau ^2/3 \right) $ and integrating over $\tau$ ,we have
\begin{equation}\label{eq:sixteen}
	\begin{aligned}
		I_e&=\frac{e^2m^2}{64\hbar ^2}\int_0^{\infty}{dv\frac{e^{-f}}{\left( 1+v \right) ^3}}fv
		\\
		&\times \left[ 3+29\left( 1+v \right) ^2-3fv\left( 2+v \right) +f^2v^2 \right]
	\end{aligned}
 \end{equation}
\begin{equation}\label{eq:seventeen}
	I_-=\frac{3e^2m^2}{16\hbar ^2}\int_0^{\infty}{dv\frac{e^{-f}}{\left( 1+v \right) ^2}}fv\left( 4+3v-fv \right) ,
\end{equation}
where $f=v\sqrt{3\left( 1+v \right)}/\chi_e $.

Finally, by summing the photon energy spectrum, one can obtain the total linear-polarization of photons $<\xi_3>$:
\begin{equation}\label{eq:eighteen}
	<\xi_3>=\frac{I_-}{I_e}=\frac{12\int_0^{\infty}{dv\frac{e^{-f}}{\left( 1+v \right) ^2}}fv\left( 4+3v-fv \right)}{\int_0^{\infty}{dv\frac{e^{-f}}{\left( 1+v \right) ^3}}fv\left[ 3+29\left( 1+v \right) ^2-3fv\left( 2+v \right) +f^2v^2 \right]}	
\end{equation}

The dependency of $<\xi_3>$ on $\chi_e$ can be calculated according to Eq.~(\ref{eq:eighteen}) and the result is shown in the Figure~\ref{fig8}.

\end{spacing}

%
\end{document}